\documentclass[prl,twocolumn,superscriptaddress,a4paper,twoside]{revtex4-1}

\usepackage{amsmath}
\usepackage{amssymb}
\usepackage{xspace}
\usepackage{graphicx}
\usepackage{grffile}
\usepackage{nicefrac}
\usepackage{tikz}
\usepackage{gensymb}
\usepackage{color}

\renewcommand{\exp}[1]{\ensuremath{\mathrm{e}^{#1}}}
\renewcommand{\vec}[1]{\ensuremath{\mathbf{#1}}}

\newlength\cstatecirclesize
\newlength\cstatetrianglelength
\newlength\cstatelinelength
\newcommand{\cstate}{\draw circle (\cstatecirclesize) ++ (0:\cstatetrianglelength) circle (\cstatecirclesize) ++ (120:\cstatetrianglelength) circle(\cstatecirclesize);}

\newcommand{\cstatethree}{\cstate\draw[red,very thick](0,0)++(180:0.5\cstatelinelength)--++(0:\cstatelinelength);}

\newcommand{\cstatesix}  {\cstate\draw[red,very thick](0:\cstatetrianglelength)++(180:0.5\cstatelinelength)--++(0:\cstatelinelength);}

\begin{document}

\title{Hidden spin-orbital hexagonal ordering induced by strong correlations in LiVS$_2$}

\author{L. Boehnke}
\affiliation{D{\'e}partement de Physique,
Universit{\'e} de Fribourg, 1700 Fribourg, Switzerland}
\affiliation{I. Institut f{\"u}r Theoretische Physik,
Universit{\"a}t Hamburg, D-20355 Hamburg, Germany}
\author{A. I. Lichtenstein}
\affiliation{I. Institut f{\"u}r Theoretische Physik,
Universit{\"a}t Hamburg, D-20355 Hamburg, Germany}
\author{M. I. Katsnelson}
\affiliation{Institute for Molecules and Materials, Radbound University Nijmegen, NL-6525 AJ Nijmegen, The Netherlands}
\author{F. Lechermann}
\affiliation{I. Institut f{\"u}r Theoretische Physik,
Universit{\"a}t Hamburg, D-20355 Hamburg, Germany}

\begin{abstract}
We present a first-principles many-body analysis of multi-orbital lattice susceptibilities in the metallic phase of the quasi-twodimensional compound LiVS$_2$. We base this on advanced correlated electronic structure methods for the $t_{2g}$-shell to reveal a highly entangled spin-orbital hexagonal ordering (SOHO) bringing about an inherently intersite order parameter for the trimerization transition, which eventually leads to an intriguing insulating phase at low temperature.
\end{abstract}
\maketitle

At low-enough temperature most solid-state systems enter a long-range ordered phase due to various instabilities. While the characterization of these ordered phases is in many cases obvious and an order parameter may be readily identified, for some materials a microscopic specification remains nebulous. Apart from the canonical magnetic, charge, orbital, structural and superconducting types of ordering more complicated phenomena have been discussed in certain materials. The famous `hidden order' state in the heavy-fermion compound URu$_2$Si$_2$~\cite{myd11,cha13,myd14} stands out as a prominent example thereof. As we will show, physics prevalent in the vanadium sulfide LiVS$_2$~\cite{kat09} can harbor hidden order with the need for a manifest multi-site description. This is due to the nature of interacting $3d$ electrons stemming from multi-orbital sites on a geometrically frustrated lattice. From model-Hamiltonian studies it is known that quantum $S=\frac{1}{2}$ spins on a frustrated lattice may give rise to many different stable phases, ranging from (anti)ferromagnets to (resonating) valence-bond solids~\cite{die05}. Adding itinerancy and further orbital differentiation entangles charge, orbital, spin as well as lattice degrees of freedom in an intriguing way and allows for unusual metallicity with proximity to unconventional ordering modes. 

The physics of the effective triangular-lattice compound LiVS$_2$~\cite{vla71} naturally addresses this sophisticated entanglement of different degrees of freedom. Within the LiVX$_2$, X=(O,S,Se) series (compare appendix~A for a qualitative phase diagram) the oxide is an insulator exhibiting an ordering transition at a critical temperature $T_\mathrm{c}\sim500$~K. The selenide has metallic character throughout the studied temperature regime. Upon cooling only the sulfide displays a metal-insulator transition (MIT) at $T_\mathrm{MIT}\sim310$~K from a paramagnetic metal to an insulating state with vanishing uniform magnetic susceptibility~\cite{kat09,kaw11}. In the insulator a spin-peierls-like lattice distortion occurs~\cite{kat09}, leading to a trimerization in the quasi-twodimensional arrangement of VS$_6$ octahedra. Orbital ordering has long been suggested to be a vital ingredient of the low-temperature phase in many layered vanadates~\cite{pen97} and experimental evidence appeared recently~\cite{mcq08,jin13}. Profound questions exist in view of the temperature depending electronic states and in general concerning the interplay of the various degrees of freedom in vicinity to the MIT in the case of LiVS$_2$. Namely the role of excitations and possible fingerprints of the ordered state within the metallic regime is of vital importance for the general understanding of metal-insulator transitions beyond the traditional Mott- and Slater-type scenarios. Moreover the VS$_2$-layer building block is also a object of interest in the context of novel time-dependent electronic structure studies~\cite{bro13}.

It is important to realize that the MIT in LiVS$_2$ may not readily be understood from a weak-coupling nesting picture. At elevated temperatures large local V moments are revealed in the paramagnetic phase of LiVO$_2$ by x-ray-absorption spectroscopy~\cite{pen98}. After their finding of anomalous metallicity above the ordering transition, Katayama {\sl et al.}~\cite{kat09} categorize the sulphur compound as a correlated paramagnetic metal at high $T$. Strong local Coulomb interactions within the nominal $3d^2$ valence of the V$^{3+}$ ion are therefore a main driving force behind the materials' phenomenology. Previous model studies for these systems from the model many-body viewpoint were based on low-order perturbation arguments~\cite{pen97}, on exact-diagonalization investigations~\cite{pen97,pen98,guo11} as well as on classical Monte-Carlo and Hartree-Fock examinations~\cite{yos11}. The first-principles electronic structure has also been addressed by density functional theory (DFT), and static strong-correlation aspects therein via the DFT+ Hubbard $U$ methodology~\cite{ezh98,guo11}. For the intricate low-temperature trimerized phase of LiVS$_2$, the model approaches favored two possible orbitally-ordered spin-state candidates, that both account for the vanishing spin susceptibility. First, a trimer singlet state connecting to the experimentally revealed dominant high-spin $S=1$ multiplet for the V ion in the metallic phase~\cite{pen97,pen98}. Second, a trimer state with low-spin $S=0$ for each V ion, i.e. violating Hund's rule~\cite{yos11}. Note that the latter proposition is different from a weak-coupling band-insulating state with unpolarized bonding orbitals. 

In order to have a full account of the realistic quantum many-body problem, we go beyond the model perspective as well as static strong-correlation investigations. We show that by lowering the temperature $T$ for LiVS$_2$ the optimal compromise between multi-orbital correlations driven by Hubbard $U$ and Hund's $J_{\rm H}$ as well as the given hopping processes on the effective triangular lattice is provided by an even more challenging ordering beyond these suggestions. Namely, by means of advanced first-principles many-body theory a spin-orbital hexagonal ordering (SOHO) is identified to originate from the metallic high-symmetry phase. Therewith for multi-orbital frustrated lattice systems a conclusive connection is drawn between a local high-spin phase at elevated $T$ and a global low-spin phase at low $T$.  


\begin{figure}
\centering
\hspace*{-0.1cm}\includegraphics{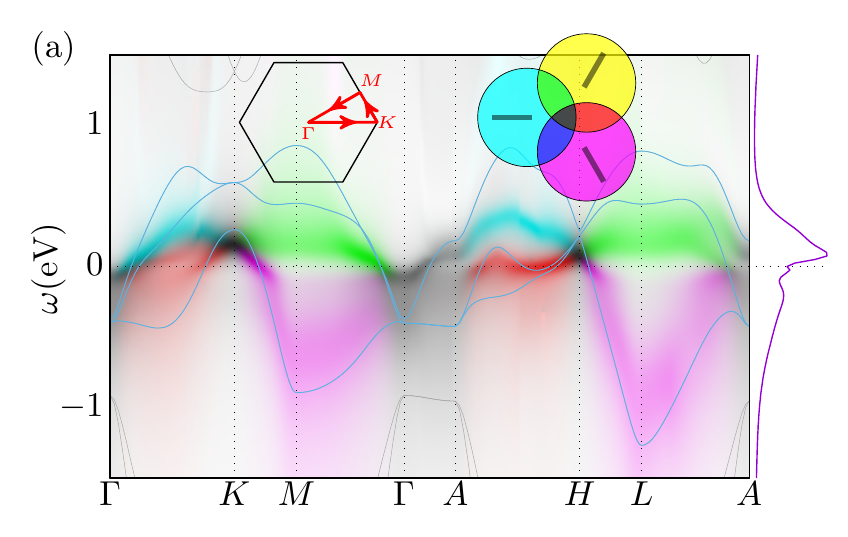}\\
\includegraphics{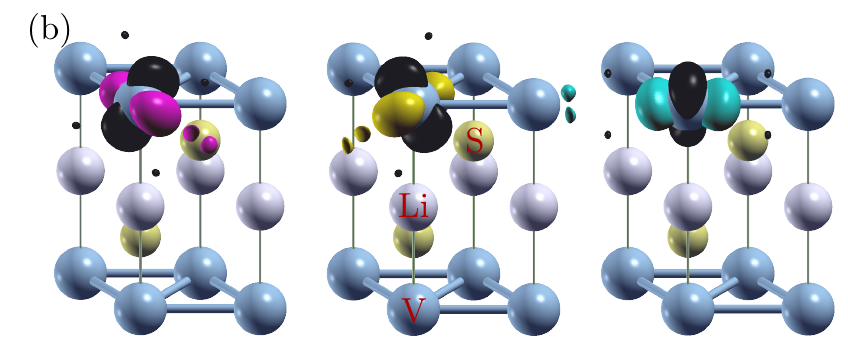}
\caption{(a) Low-energy interacting DFT+DMFT spectral function of LiVS$_2$ slightly above the transition temperature with comparison to the DFT bands (full lines) shown along the path through the Brillouin zone noted in the left inset. The represented spectral weight differentiates between the three correlated-subspace contributions using an subtractive color scheme based on the color coding for the Wannier functions, which is shown in the right inset. (b) Low-energy degenerate $t_{2g}$-like Wannier functions (with specific color coding: magenta, yellow, cyan).\label{fig:spec}}
\end{figure}
Details of the density functional and dynamical mean-field theory (DFT+DMFT)\cite{geo96,ani97,lic98,kot06} calculations are given in appendix~B. Figure~\ref{fig:spec} displays band-narrowing, transfer of spectral weight as well as lifetime effects in the one-particle spectral function $A(\vec{k},\omega)$ due to the rotational-invariant multi-orbital Coulomb interactions on each V site. Especially along $\Gamma K$ and the corresponding $AH$ direction at $k_z=\pi$ in the Brillouin zone (BZ) the renormalization leads to intricate very low-energy susceptible many-body states close to $\varepsilon_{\rm F}$. Notably the electron pocket at $\Gamma$ and the hole pocket at $K$ get shifted towards the Fermi level. The Wannier function-character contribution varies strongly in the inplane $\vec{k}$ directions and band-like coherency is quickly lost away from the Fermi level, specifically in the unoccupied higher-energy region.


DFT+DMFT describes a strongly correlated metal at elevated temperatures. A large local magnetic moment on the V ions associated with $\langle S_{\rm loc}^2\rangle\sim1.96$ is retrieved. As outlined, by lowering $T$ an intriguing MIT scenario sets in that involves nearly all available system degrees of freedom. The remaining challenge to shed light on the LiVS$_2$ ordering relies in the combination of nonlocal physics, i.e. real-space order parametrization that involves correlations among different lattice sites, with manifest multi-orbital degrees of freedom. To tackle this we advance the DFT+DMFT approach by appending a two-particle-susceptibility formalism that includes generic multi-orbital vertex contributions. This allows to study quantum fluctuations leading to nonlocal ordering tendencies in the correlated metallic high-temperature regime above $T_{\rm MIT}$, in principle without breaking translational symmetry in real- or reciprocal space. Hence instead of addressing the broken-symmetry phase directly, we remain in the metallic state and examine multi-orbital two-particle response functions upon lowering the temperature. That approach is indeed adequate in the present context, since diffuse scattering hinting towards precursive manifestations of the ordered state has been noticed in the electron-diffraction pattern of metallic LiVS$_2$~\cite{kat09}.

In general, phase transitions are indicated by a divergence of the static susceptibility associated with the underlying order parameter and with a wave vector signaling the real-space pattern of the ordered phase. Beyond former single-band studies~\cite{mai05,boe12}, our approach allows access to the complete three-orbital particle-hole susceptibility tensor $\chi^{\sigma\sigma'}_{mm'm''m'''}(\vec{q},\omega)$ at finite temperature, with full generality concerning its frequency-dependent structure~\cite{boe11}. It allows an evaluation of all experimentally measurable susceptibilities and even explicit determination of the order parameter.\cite{boe15}

In the case of models with a single correlated orbital per site, the longitudinal particle-hole channel allows for 2 susceptibilities, namely the ($\vec{q}$-dependent) spin- and charge response~\cite{boe11}. For a three-orbital $t_{2g}$ shell however, there are 18 such independent possible susceptibilities and not much problem-tailored physical insight may be gained by monitoring all of those. A much more promising route to examine susceptibilities in multi-orbital materials is to focus in a first step on the eigenvalues/modes of the susceptibility tensor $\chi_{\alpha\beta}$ in the product basis $\alpha=\{\sigma mm'\}$ and $\beta=\{\sigma' m'''m''\}$ with $m,m'm'',m'''$=\tikz[baseline=-3pt]{\draw[magenta,very thick](0,0)++(120:0.2cm)--++(300:0.4cm);},\tikz[baseline=-3pt]{\draw[yellow!80!black,very thick](0,0)++(60:0.2cm)--++(240:0.4cm);},\tikz[baseline=-3pt]{\draw[cyan,very thick](0,0)++(0:0.2cm)--++(180:0.4cm);} and $\sigma,\sigma'=\uparrow,\downarrow$.
From such an analysis
\begin{align}
  \label{eqn:eigendecomposition}
  \chi^{(l)}(\vec{q},\mbox{$\omega$=0})=&\left<\mathcal{T}_\tau\sum_\alpha \widehat{v}^{(l)}_\alpha(\vec{q}) \sum_\beta\widehat{v}^\star{}^{(l)}_\beta(\vec{q})\right>\\
=&\left<\mathcal{T}_\tau\widehat{V}^{(l)}(\vec{q})\widehat{V}^\star{}^{(l)}(\vec{q})\right>
\end{align}
is the $l$th eigenvalue and $\widehat{V}^{(l)}=\sum_{\sigma mm'}v^{(l)}_{\sigma mm'}c^\dagger{}^\sigma_mc^\sigma_{m'}$ is the corresponding eigenmode ($v^{(l)}_\alpha$ is the $l$th eigenvector of the susceptibility tensor $\chi_{\alpha\beta}$). This makes $\widehat{V}^\mathrm{max}$ the dominant fluctuating excitation upon approaching the phase transition and $\theta=\nobreak \big<\widehat{V}^{\mathrm{max}}\big>$ a natural order parameter for the transition.

\begin{figure}
\centering
  \includegraphics{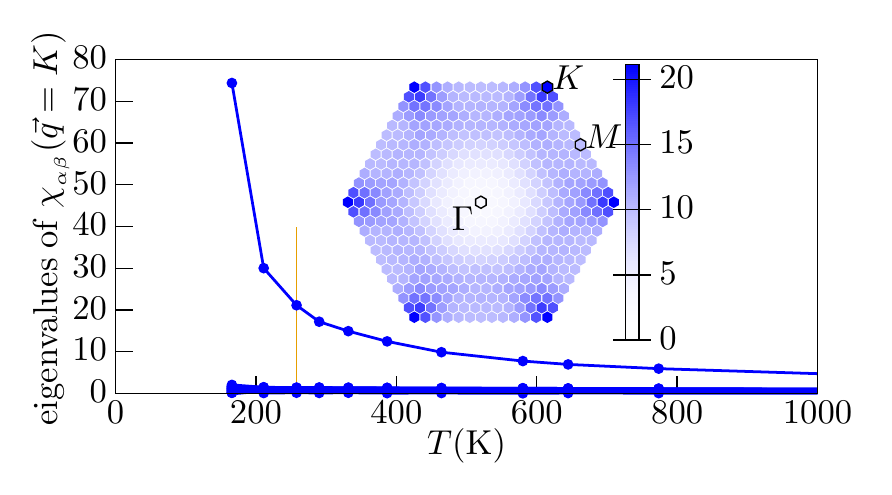}
\caption{Temperature dependence of all particle-hole-based susceptibility eigenvalues at the $K$ point. The inset shows the largest eigenvalue throughout the Brillouin zone, emphasizing that relevant susceptibilities reside at the $K$ point. The vertical line marks $T=260$~K as the temperature used for the inset and for all other calculations throughout the work, if not stated otherwise.}
\label{fig:chivsT}
\end{figure}
Figure~\ref{fig:chivsT} shows the eigenvalue-evolution of $\chi_{\alpha\beta}(K,\mbox{$\omega$=0})$ with temperature. The $K$ point being the position of the prominent maximum, as visible in the inset of Fig.~\ref{fig:chivsT}. A single eigenvalue diverges, indicating a phase transition at roughly $T_{\rm crit}\sim150$~K, undershooting the experimentally observed transition temperature $T_{\rm MIT}\sim310$~K. However this is to be expected~\cite{pav08}, given the fact that our model neglects phonon contributions and thus does not take into account lattice distortions that accompany the transition~\cite{kat09}.

\begin{figure}
  \centering
  \includegraphics{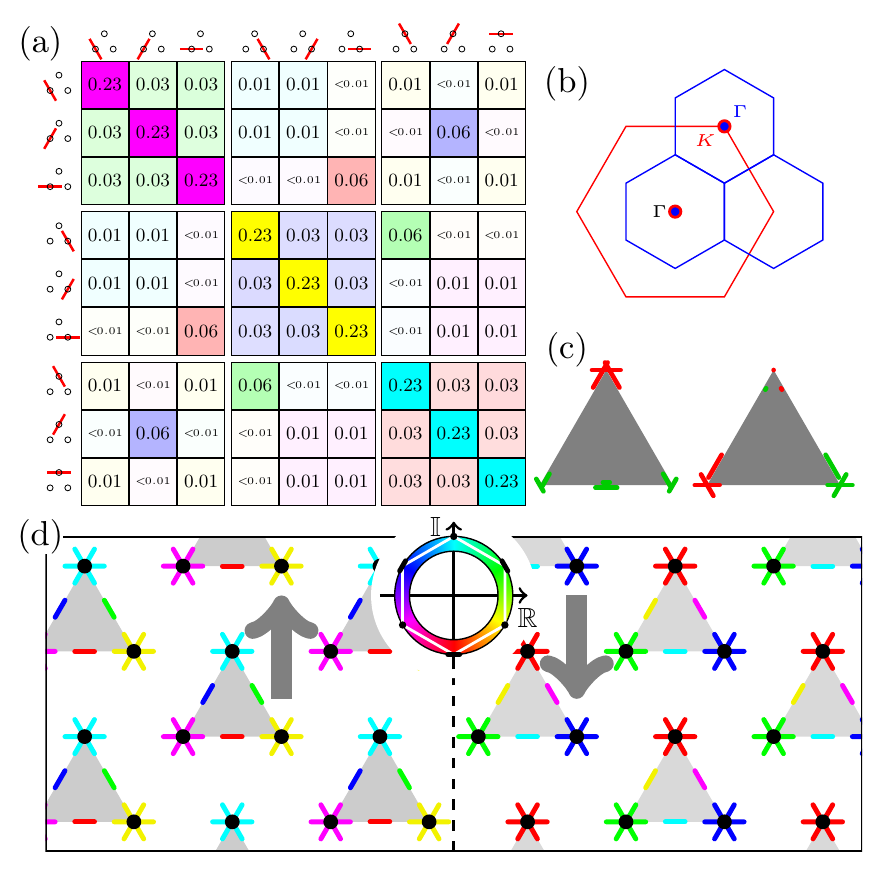}
  \caption{(a)  Site-orbital resolved maximum eigenmode contributions. Only the $\uparrow$-part $v^{\mathrm{max}}_{\uparrow mm'}$ is shown with $m$ and $m'$ the adumbrated orbital indices on the supercell. (b) Original (red) and super-lattice (blue). The original lattice $K$ point coincides with the superlattices $\Gamma$ point. (c) Possible imprinted spin orderings on the triangular building block. (d) Identified LiVS$_2$ order at low temperature by visualizing the maximum spin-orbital resolved eigenmode contribution $v^{\mathrm{max}}_{\sigma mm'}$ of a. The inset shows the colorcoding of the complex values used in a and d. Lines highlight the hexagonal phase relation between the major contributions to the order parameter.}
  \label{fig:order}
\end{figure}

The above approach assumes a spatial variation of an single-site order parameter, which when reviewing experimental findings~\cite{kat09} or the anisotropy of the three respective orbitals (compare Fig.~\ref{fig:spec}b) can not be taken for granted. Yet, it gives the right hint towards a spin-like ($\widehat{V}^\mathrm{(max)}(K)\propto\widehat{S}^\mathrm{(tot)}_z$) K point ($\sqrt{3}\times\sqrt{3}$ superlattice) ordering instability. Having uncovered this pattern, in a second step we expand the multi-orbital particle-hole susceptibility investigation by deliberately breaking translational symmetry of the lattice two-particle Green's function in this way. Thereto, we solve the supercell (SC) Bethe-Salpeter equation (BSE)
\begin{equation}
  \label{eqn:BSE}
 \widetilde{\chi}^{-1}_\mathrm{SC}(\vec{Q})=\widetilde{\chi}^{(0)}_\mathrm{SC}{}^{-1}(\vec{Q})+\gamma_\mathrm{SC}
\end{equation}
on the LiVS$_2$ triangular lattice build from a minimal triangle three-site basis with superlattice wave vector $\vec{Q}=\Gamma$ (compare Fig.~\ref{fig:order}c). All quantities in eq.~(\ref{eqn:BSE}) carry the full inner fermionic degrees of freedom, with a Legendre representation replacing fermionic Matsubara frequencies~\cite{boe11} which renders this calculation both accurate and numerically possible. The undistorted one-particle Green's function is used for the bare polarization $\widetilde{\chi}^{(0)}$ to take into account that our investigations still deal with temperatures \emph{above} the transition to the disordered phase. The constructed irreducible supercell vertex function $\gamma_c$ remains site-diagonal. That way, we anticipate the translational symmetry breaking of the high-temperature susceptibility shown in Fig.~\ref{fig:chivsT}.

Analysis of the 162 eigenmodes of the resulting supercell susceptibility tensor reveals two degenerate dominant eigenvalues, being partners in a two-dimensional irreducible representation of the triangular building block. These eigenmodes are given Fig.~\ref{fig:order}a, for convenience rotated into the most prudent basis of the two-dimensional eigenspace, diagonalizing the 120\degree{} rotation, which results in a complex order parameter. Only one of the two partners is shown, the other is its complex conjugate. Also, $v^{\mathrm{max}}_{\downarrow mm'}=-v^{\mathrm{max}}_{\uparrow mm'}$ as is typical for spin-like excitations. The interesting feature is the intricate orbital degree of freedom. The large diagonal parts with the $\exp{i\frac{2}{3}\pi}$, i.e. 120$^{\circ}$-degree, phase shift between lattice sites is reminiscent of a $K$-point excitation on a triangular lattice. The same pattern can be observed in models without orbital anisotropy. On the other hand, the appearance of relevant inter-site elements reflects the building of a triangular supercell molecular orbital on the periodic supercell as sketched in Fig.~\ref{fig:order}a. A key feature of this ordering mode is the phase relation between the different onsite and intersite spin-orbital parts on the elementary triangle. As seen in Fig.~\ref{fig:order}a,b, the respective phases form a hexagon in the complex plane and hence the overall state may be accurately refered to as spin-orbital hexagonal order (SOHO).

This traceless hidden-excitation mode does not imprint a finite magnetic moment to the supercell. Also applying a magnetic field to this robust system does not easily affect the total supercell spin arrangement. That renders our proposed order consistent with the observation of dramatic decrease of uniform magnetic susceptibility in the ordered phase~\cite{kat09} despite the large local magnetic moment in the disordered phase. The large vanadium local moment at elevated temperature becomes locked in a valence-bond-like fashion within the insulator, therefore not allowing for couplings to moderate applied magnetic fields. It is important to explicitly break the translational symmetry within the BSE. An eigenmode analysis of a three-site supercell susceptibility obtained from a single-site BSE does not reveal the non-site-diagonal contributions.

As we concentrate on the structurally undistorted lattice, the observed fluctuating order maintains equal occupations of the orbitals. Orbital-symmetry breaking follows already on the DFT level (see appendix~C for detailed discussion) as soon as the trimerization and the accompanying lattice distortion set in. At that point, we also observe a difference in occupation, favoring the orbitals with a neighbor within the triangle. The weakly-occupied orbital on each V atom in the insulating phase does not have relevant non site-diagonal elements for the order parameter (see Fig.~\ref{fig:order}a). Thus it does not participate in the build-up of supercell molecular orbitals and we do not expect a relevant alteration from the fluctuating excitation to the realized broken translational-symmetry state.

The order parameter in Fig~\ref{fig:order}a does not readily provide an easy physical interpretation. To that end, we slightly simplify the order parameter by keeping only the twelve largest contributions, which allows for an easy interpretation of the dominant effect (see appendix~D for more details). Fig~\ref{fig:order}b gives a visualization of the spin-imprint on the triangular building block, where red/green correspond to $\uparrow$/$\downarrow$ spins, and lines along a side of the triangle emblematize superpositions of the aligned orbitals on the adjacent sites. Two representatives of the space of easily excitable states are choosen. Whether this space reduces to isolated solutions upon explicit breaking of the translational symmetry is an interesting question to be investigated.

After the characterization of the obtained complex ordering, we also want to touch base with the ordered-state propositions from previous works on the LiVX$_2$, (X=O,S) compounds. The present findings for the order parameter of the trimerization phase transition enables us to assess the overlap with other suggested ground states $\left|\psi\right>$ by readily evaluating the ordering amplitude $\theta_{\rm trial}=\big<\psi\big|\widehat{V}^\mathrm{max}\big|\psi\big>$. Specifically, the onsite low-spin scenario proposed by Yoshitake {\it et al.}~\cite{yos11} evaluates to zero, thus does not constitute a viable option. This is not surprising, because a key ingredient of our ordering mode is the intact local $S=1$ vanadium spin. On the other hand, Pen {\it et al.}~\cite{pen98} proposed an onsite high-spin ordered state in the form of a product wave function of three local triplets on each equilateral triangle of V($3d^2$) ions. Indeed such a state yields a non-zero expectation value for the ordering amplitude $\theta_{\rm trial}$. But it does so solely based on the diagonal elements $v^\mathrm{max}_{\sigma mm}$ of the dominant fluctuating superlattice excitation, in line with the purely local picture. The relevant intersite terms are missing and thus that state serves only as an approximant to the true more complicate order. In the model picturing of~\cite{pen98} a strong Hund's $J_{\rm H}$ seems to overrule the nearest-neighbor intersite exchange $J_{\rm NN}$ (here not simply assumed to be in the Heisenberg-limit of the Hubbard model) in a complete fashion. Whereas our findings reveal a subtle interplay between local and nonlocal exchange processes that give rise to substantial interorbital intersite terms in the description of the ordered state.

The present work enriches the plethora of categories in the physics of metal-insulator transitions by revealing the challenging connection between the high-temperature metallic phase with large magnetic susceptibility and the low-temperature ordered phase with zero magnetic susceptibility in a frustrated multi-orbital compound subject to strong correlations. To this end, a powerful first-principles many-body analysis of multi-orbital lattice susceptibilities in the disordered state is introduced to investigate phase transitions in correlated materials. It complements existing approaches employed directly in the ordered phase and may be applied to further solid-state problems of strong correlations. For LiVS$_2$ a complex SOHO mode originating in the metal leads to a trimerized insulator with a unique electronic structure beyond standard Mott-insulating mechanisms. The identified phase-sensitive $S^{(\rm tot)}_z=0$ ordered state generalizes the valence-bond concept of single-orbital $S=\frac{1}{2}$ systems to multi-orbital $S=1$ problems on a frustrated lattice. Electron- or hole doping of that new state is believed to lead to fascinating metallicity with unique transport properties. Moreover investigating LiVS$_2$ under pressure or applying directional strain most likely results in emerging net-moment magnetism due to unlocking of spins on the equilateral base structure.

\begin{acknowledgments}
We thank H. Takagi for helpful discussions. The work benefits from financial support through the DFG-FOR1346 and the DFG-SFB925. Calculations were performed at the North-German Supercomputing Alliance (HLRN).
\end{acknowledgments}
\appendix
\begin{figure}
\centering
  \includegraphics{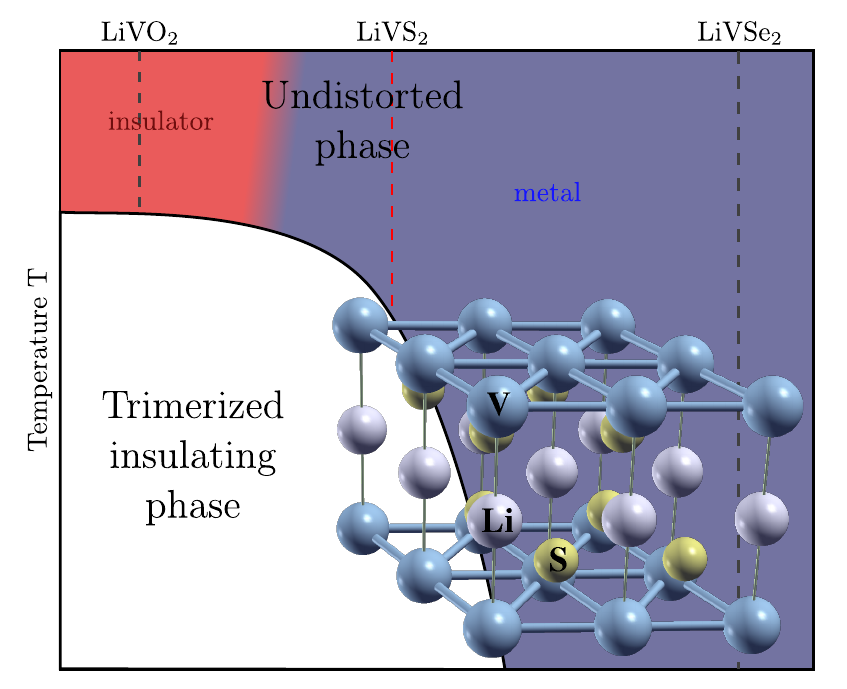}
\caption{Sketch of the generic LiVX$_2$, X=(O,S,Se) phase diagram (based on Ref.~\onlinecite{kat09}) with the P$\bar{3}$m1-LiVS$_2$ crystal structure in the metallic state as an inset.\label{fig:phases}}
\end{figure}
\section{APPENDIX A: LiVS$_2$ structural properties}
The LiVS$_2$ compound has the underlying P$\bar{3}$m1 space group ($a$=3.38 \AA, $c/a$=1.82)~\cite{vla71} with an 1$T$-type ion-stacking along the $c$-axis, the sandwich VS$_2$ layer has ideal triangular symmetry. Figure~\ref{fig:phases} gives a rough view on the phases depending on the chalcogen in the system.

\section{APPENDIX B: Details of the modeling}
Density Functional theory~(DFT) calculations in the Local Density Approximation~(LDA) implemented within a mixed-basis pseudopotential code~\cite{mbpp} and using crystal-structure data from experiment~\cite{vla71} place the Fermi level $\varepsilon_{\rm F}$ within the isolated three-band manifold of dominant $3d(t_{2g})$ character and bandwidth $W\sim2.1$~eV (compare Fig.~1a of the main text). These bands host the two electrons of the low-spin V($3d^2$) filling. For this threefold a maximally-localized Wannier-function (WF) basis~\cite{mar97,mos08} may be derived with the low-energy WFs directed along the canonical directions towards neighboring vanadium ions (see Fig~1b of the main text). The nearest-neighbor (NN) hopping within this degenerate basis amounts to $t_{\rm NN}=-290$~meV along the facing orbitals of the respective axes.

To include realistic electronic correlations beyond the static limit we continue the investigation in the state-of-the-art Density Functional and Dynamical Mean Field Theory (DFT+DMFT) framework~\cite{geo96,ani97,lic98,kot06} by utilizing the derived WF basis as the correlated subspace. The resulting orbital-dependent strong electronic correlations significantly modify the LDA-derived electronic structure (Fig.~1a of the main text). Interestingly, since there are two electrons in three orbitals, LiVS$_2$ falls in the category of an Hund's metal where besides the Hubbard $U$ the local interorbital (Hund's) exchange $J_{\rm H}$ has a dominant influence on the strong-correlation physics (see e.g.~\cite{geo13} for a recent review).

For the DMFT part an hybridization-expansion continuous-time (CT-Hyb)~\cite{wer06,gul11} Quantum Monte Carlo (QMC) impurity solver as implemented in the~\textsc{TRIQS} package~\cite{par15,set15} is utilized. Therein advantage is taken of the Legendre representation of the Green's function~\cite{boe11}. We consider the most general form for the rotational invariant Coulomb interaction~\cite{str14}, restricted to the $t_{2g}$ subspace, where it takes the Slater-Kanamori form~\cite{dem11}
\begin{equation}
\hat{H}_{\mathrm{int}}=(U-3J_\mathrm{H})\frac{\hat{N}(\hat{N}-1)}{2}-2J_\mathrm{H}\vec{S}^2-\frac{J_\mathrm{H}}{2}\vec{L}^2+\frac{5}{2}\hat{N}\\
\end{equation}
with $\hat{N}$ is the total charge operator, $\vec{S}$ the spin- and $\vec{L}$ the angular momentum operators. We chose $U=3.5$~eV and $J_\mathrm{H}=0.7$~eV, appropriate for vanadium sulfides~\cite{gri10}.

At DMFT self-consistency, also the full multi-orbital susceptibility tensor is evaluated in a DMFT-like approximation, e.g. assuming the locality of the two-particle particle-hole irreducible vertex in the Bethe-Salpeter equation (BSE) in this channel, as valid in the infinite-dimension limit~\cite{zla90}. This approximation has successfully been used for (effective) single-orbital problems~\cite{mai05,boe12}. Note that already on the latter level the calculations are numerically demanding. The inversion of the BSE requires the Monte-Carlo accumulation and the handling of the two-particle Green's function with its four orbital dependencies and full account of also the inner fermionic Matsubara frequencies. For the latter we employ the Legendre representation~\cite{boe11}, as its better convergence helps tremendously in solving the superlattice BSE (eq.~(3) in the main text).

In the BSE, the longitudinal and the transversal susceptibilities are decoupled, allowing a sole investigation of the longitudinal components (orbitally resolved $\hat{S}_z$ and $\hat{n}$-like excitations).

\section{APPENDIX C: The distorted lattice problem}
\begin{figure}
\centering
  \includegraphics{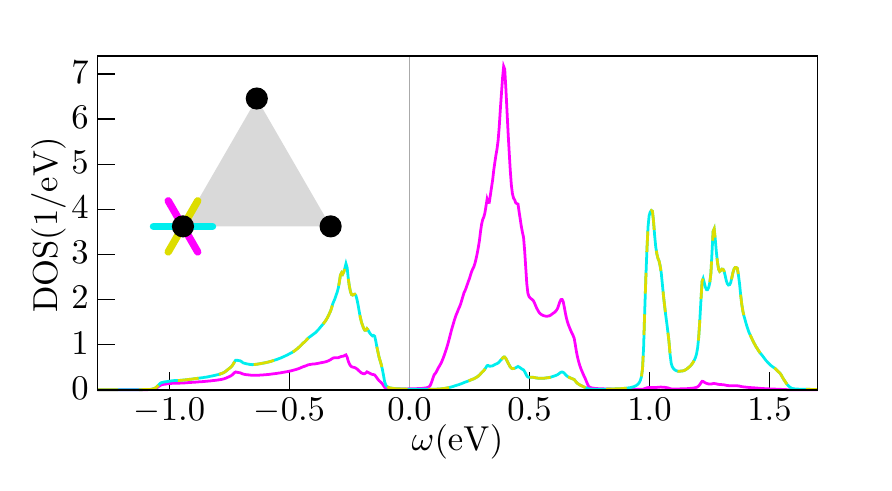}
\caption{Local $t_{2g}$ LDA density of states of the insulating structurally distorted low-temperature compound. Local orbital ordering via the dominant occupation of the yellow and the cyan orbital (for the choice of the V site adumbrated in the inset) is observable.
\label{fig:insulator}}
\end{figure}
LDA calculations in the low-temperature $\sqrt{3}\times\sqrt{3}$ phase with the experimentally observed lattice distortion on the quasi-twodimensional triangular VS$_2$ lattice~\cite{kaw11} indeed reveal an insulating state (see Fig.~\ref{fig:insulator}). This nonmagnetic band insulator in LDA displays a band gap of the order of 200~meV. On each V site of the isolated triangles, respectively, the $t_{2g}$ orbitals pointing along the triangle edges are most strongly occupied with a difference in occupation $n_{\tikz[baseline=-3pt]{\draw[yellow!80!black,very thick](0,0)++(60:0.1cm)--++(240:0.2cm);}}=n_{\tikz[baseline=-3pt]{\draw[cyan,very thick](0,0)++(0:0.1cm)--++(180:0.2cm);}}\approx0.83$, $n_{\tikz[baseline=-3pt]{\draw[magenta,very thick](0,0)++(120:0.1cm)--++(300:0.2cm);}}\approx0.35$. However this effective single-particle band-insulating solution is motivated from the sole observed symmetry reduction from the structural distortion. It cannot give a full picture, since e.g. the sophisticated fate of the high-$T$ paramagnetic local spin degree of freedom close to and in the ordered phase is left unexplained from this simplistic viewpoint and thus fails to reveal the electronic mechanism behind the trimerization.

\section{APPENDIX D: The fluctuation eigenbasis}
The main manuscript gives a representation of the fluctuation in terms of the $t_{2g}$ states of the supercell sites (figure 3c of that publication). The structure of this has been discussed, most prominently the relevant offdiagonal components connecting facing orbitals between different sites of the supercell.

Going beyond the general remark that these components describe cluster-states as opposed to site-states, this section will give a quantitative, yet approximate, evaluation of the eigenbasis of the dominant fluctuating mode.

To ease the interpretation of this eigenbasis, it is convenient to restrict the matrix in Fig.~3a of the main text to the diagonal elements and the six largest non-diagonal ones. This discards site-diagonal orbital-offdiagonal elements, which would obfuscate the relevant structure of eigenstates of the dominant fluctuating mode by introducing small admixtures of the different orbitals of the same site.

With those aside, the task comes down to finding the eigenbases of the real and imaginary part of the complex eigenmode
\begin{widetext}
\begin{align}
  \Re S^\mathrm{max}_{mm'}=&\Re(V^\mathrm{max}_{\uparrow mm'}-V^\mathrm{max}_{\downarrow mm'})\label{eqn:livs2realbasis}
  =\small
\begin{pmatrix}
          - 0.40 &  \phantom{-0.0}0 &  \phantom{-0.0}0 &  \phantom{-0.0}0 &  \phantom{-0.0}0 &  \phantom{-0.0}0 &  \phantom{-0.0}0 &  \phantom{-0.0}0 &  \phantom{-0.0}0 \\
 \phantom{-0.0}0 &           - 0.40 &  \phantom{-0.0}0 &  \phantom{-0.0}0 &  \phantom{-0.0}0 &  \phantom{-0.0}0 &  \phantom{-0.0}0 &           - 0.10 &  \phantom{-0.0}0 \\
 \phantom{-0.0}0 &  \phantom{-0.0}0 &           - 0.40 &  \phantom{-0.0}0 &  \phantom{-0.0}0 &  \phantom{-0.0}0 &  \phantom{-0.0}0 &  \phantom{-0.0}0 &  \phantom{-0.0}0 \\
 \phantom{-0.0}0 &  \phantom{-0.0}0 &  \phantom{-0.0}0 & \phantom{-} 0.40 &  \phantom{-0.0}0 &  \phantom{-0.0}0 & \phantom{-} 0.10 &  \phantom{-0.0}0 &  \phantom{-0.0}0 \\
 \phantom{-0.0}0 &  \phantom{-0.0}0 &  \phantom{-0.0}0 &  \phantom{-0.0}0 & \phantom{-} 0.40 &  \phantom{-0.0}0 &  \phantom{-0.0}0 &  \phantom{-0.0}0 &  \phantom{-0.0}0 \\
 \phantom{-0.0}0 &  \phantom{-0.0}0 &  \phantom{-0.0}0 &  \phantom{-0.0}0 &  \phantom{-0.0}0 & \phantom{-} 0.40 &  \phantom{-0.0}0 &  \phantom{-0.0}0 &  \phantom{-0.0}0 \\
 \phantom{-0.0}0 &  \phantom{-0.0}0 &  \phantom{-0.0}0 & \phantom{-} 0.10 &  \phantom{-0.0}0 &  \phantom{-0.0}0 &  \phantom{-0.0}0 &  \phantom{-0.0}0 &  \phantom{-0.0}0 \\
 \phantom{-0.0}0 &           - 0.10 &  \phantom{-0.0}0 &  \phantom{-0.0}0 &  \phantom{-0.0}0 &  \phantom{-0.0}0 &  \phantom{-0.0}0 &  \phantom{-0.0}0 &  \phantom{-0.0}0 \\
 \phantom{-0.0}0 &  \phantom{-0.0}0 &  \phantom{-0.0}0 &  \phantom{-0.0}0 &  \phantom{-0.0}0 &  \phantom{-0.0}0 &  \phantom{-0.0}0 &  \phantom{-0.0}0 &  \phantom{-0.0}0 
\end{pmatrix}
\end{align}
\begin{align}
  \Im S^\mathrm{max}_{mm'}=&\Im(V^\mathrm{max}_{\uparrow mm'}-V^\mathrm{max}_{\downarrow mm'})\label{eqn:livs2imagbasis}
  =\small
\begin{pmatrix}
          - 0.23 &  \phantom{-0.0}0 &  \phantom{-0.0}0 &  \phantom{-0.0}0 &  \phantom{-0.0}0 &  \phantom{-0.0}0 &  \phantom{-0.0}0 &  \phantom{-0.0}0 &  \phantom{-0.0}0 \\
 \phantom{-0.0}0 &           - 0.23 &  \phantom{-0.0}0 &  \phantom{-0.0}0 &  \phantom{-0.0}0 &  \phantom{-0.0}0 &  \phantom{-0.0}0 & \phantom{-} 0.06 &  \phantom{-0.0}0 \\
 \phantom{-0.0}0 &  \phantom{-0.0}0 &           - 0.23 &  \phantom{-0.0}0 &  \phantom{-0.0}0 &           - 0.12 &  \phantom{-0.0}0 &  \phantom{-0.0}0 &  \phantom{-0.0}0 \\
 \phantom{-0.0}0 &  \phantom{-0.0}0 &  \phantom{-0.0}0 &           - 0.23 &  \phantom{-0.0}0 &  \phantom{-0.0}0 & \phantom{-} 0.06 &  \phantom{-0.0}0 &  \phantom{-0.0}0 \\
 \phantom{-0.0}0 &  \phantom{-0.0}0 &  \phantom{-0.0}0 &  \phantom{-0.0}0 &           - 0.23 &  \phantom{-0.0}0 &  \phantom{-0.0}0 &  \phantom{-0.0}0 &  \phantom{-0.0}0 \\
 \phantom{-0.0}0 &  \phantom{-0.0}0 &           - 0.12 &  \phantom{-0.0}0 &  \phantom{-0.0}0 &           - 0.23 &  \phantom{-0.0}0 &  \phantom{-0.0}0 &  \phantom{-0.0}0 \\
 \phantom{-0.0}0 &  \phantom{-0.0}0 &  \phantom{-0.0}0 & \phantom{-} 0.06 &  \phantom{-0.0}0 &  \phantom{-0.0}0 & \phantom{-} 0.46 &  \phantom{-0.0}0 &  \phantom{-0.0}0 \\
 \phantom{-0.0}0 & \phantom{-} 0.06 &  \phantom{-0.0}0 &  \phantom{-0.0}0 &  \phantom{-0.0}0 &  \phantom{-0.0}0 &  \phantom{-0.0}0 & \phantom{-} 0.46 &  \phantom{-0.0}0 \\
 \phantom{-0.0}0 &  \phantom{-0.0}0 &  \phantom{-0.0}0 &  \phantom{-0.0}0 &  \phantom{-0.0}0 &  \phantom{-0.0}0 &  \phantom{-0.0}0 &  \phantom{-0.0}0 & \phantom{-} 0.46 
\end{pmatrix}
\end{align}
\end{widetext}
which leads to only eigenvectors of the simple form
\begin{equation}
  \cos(\alpha)\tikz[baseline=0pt]\cstatethree;+\sin(\alpha)\tikz[baseline=0pt]\cstatesix;
\end{equation}
and of course rotated versions of those, which leads to the depiction in Fig.~3c of the main manuscript.

Each line in that figures represents one eigenvector, encoding all its information. The orientation of the line obviously identifies the two facing orbitals, the position on the edge of the triangle encodes the phase of the superposition, going from $\alpha=0$ if full relative weight is on one of the orbitals to $\alpha=\frac{\pi}{2}$ for the other.

If the constituents enter with equal sign, thus forming an even-like superposition, the line is shifted to the outside of the triangle, in the other case, for an odd-like superposition, the line is shifted to the inside of the triangle. Eigenvectors that are formed from just one site orbital are not shifted. Finally, the length of the line indicates the eigenvalue corresponding to the eigenvector, green for a positive eigenvalue, red for a negative one.

The `outer' orbitals that do not have a facing neighbor within the supercell only ever appear isolated, a property that is also obvious from just inspecting \eqref{eqn:livs2realbasis} and \eqref{eqn:livs2imagbasis} since they do not couple to other orbitals anymore in this approximation.

While merely illustrative due to its approximative nature, Fig.~3c of the main text helps to clarify the notion of the formation of cluster molecular orbital like states for the dominant order fluctuating mode. It becomes clear that while some states are still site-like, the three-site order parameter carries inherently nonlocal contributions. It shows the eigenbasis of the real- and imaginary part of the complex eigenmode, therewith of two representatives of the manifold of easily excitable (and eventually self-exciting) states, it is showing the spin-imprint of the excitation, being periodically continued on the supercluster lattice.

\bibliography{references}

\begin{thebibliography}{36}%
\makeatletter
\providecommand \@ifxundefined [1]{%
 \@ifx{#1\undefined}
}%
\providecommand \@ifnum [1]{%
 \ifnum #1\expandafter \@firstoftwo
 \else \expandafter \@secondoftwo
 \fi
}%
\providecommand \@ifx [1]{%
 \ifx #1\expandafter \@firstoftwo
 \else \expandafter \@secondoftwo
 \fi
}%
\providecommand \natexlab [1]{#1}%
\providecommand \enquote  [1]{``#1''}%
\providecommand \bibnamefont  [1]{#1}%
\providecommand \bibfnamefont [1]{#1}%
\providecommand \citenamefont [1]{#1}%
\providecommand \href@noop [0]{\@secondoftwo}%
\providecommand \href [0]{\begingroup \@sanitize@url \@href}%
\providecommand \@href[1]{\@@startlink{#1}\@@href}%
\providecommand \@@href[1]{\endgroup#1\@@endlink}%
\providecommand \@sanitize@url [0]{\catcode `\\12\catcode `\$12\catcode
  `\&12\catcode `\#12\catcode `\^12\catcode `\_12\catcode `\%12\relax}%
\providecommand \@@startlink[1]{}%
\providecommand \@@endlink[0]{}%
\providecommand \url  [0]{\begingroup\@sanitize@url \@url }%
\providecommand \@url [1]{\endgroup\@href {#1}{\urlprefix }}%
\providecommand \urlprefix  [0]{URL }%
\providecommand \Eprint [0]{\href }%
\providecommand \doibase [0]{http://dx.doi.org/}%
\providecommand \selectlanguage [0]{\@gobble}%
\providecommand \bibinfo  [0]{\@secondoftwo}%
\providecommand \bibfield  [0]{\@secondoftwo}%
\providecommand \translation [1]{[#1]}%
\providecommand \BibitemOpen [0]{}%
\providecommand \bibitemStop [0]{}%
\providecommand \bibitemNoStop [0]{.\EOS\space}%
\providecommand \EOS [0]{\spacefactor3000\relax}%
\providecommand \BibitemShut  [1]{\csname bibitem#1\endcsname}%
\let\auto@bib@innerbib\@empty
\bibitem [{\citenamefont {Mydosh}\ and\ \citenamefont
  {Oppeneer}(2011)}]{myd11}%
  \BibitemOpen
  \bibfield  {author} {\bibinfo {author} {\bibfnamefont {J.~A.}\ \bibnamefont
  {Mydosh}}\ and\ \bibinfo {author} {\bibfnamefont {P.~M.}\ \bibnamefont
  {Oppeneer}},\ }\href {\doibase 10.1103/RevModPhys.83.1301} {\bibfield
  {journal} {\bibinfo  {journal} {Rev. Mod. Phys.}\ }\textbf {\bibinfo {volume}
  {83}},\ \bibinfo {pages} {1301} (\bibinfo {year} {2011})}\BibitemShut
  {NoStop}%
\bibitem [{\citenamefont {Chandra}\ \emph {et~al.}(2013)\citenamefont
  {Chandra}, \citenamefont {Coleman},\ and\ \citenamefont {Flint}}]{cha13}%
  \BibitemOpen
  \bibfield  {author} {\bibinfo {author} {\bibfnamefont {P.}~\bibnamefont
  {Chandra}}, \bibinfo {author} {\bibfnamefont {P.}~\bibnamefont {Coleman}}, \
  and\ \bibinfo {author} {\bibfnamefont {R.}~\bibnamefont {Flint}},\ }\href
  {\doibase 10.1038/nature11820} {\bibfield  {journal} {\bibinfo  {journal}
  {Nature}\ }\textbf {\bibinfo {volume} {493}},\ \bibinfo {pages} {621}
  (\bibinfo {year} {2013})}\BibitemShut {NoStop}%
\bibitem [{\citenamefont {Mydosh}\ and\ \citenamefont
  {Oppeneer}(2014)}]{myd14}%
  \BibitemOpen
  \bibfield  {author} {\bibinfo {author} {\bibfnamefont {J.}~\bibnamefont
  {Mydosh}}\ and\ \bibinfo {author} {\bibfnamefont {P.}~\bibnamefont
  {Oppeneer}},\ }\href {\doibase 10.1080/14786435.2014.916428} {\bibfield
  {journal} {\bibinfo  {journal} {Philosophical Magazine}\ }\textbf {\bibinfo
  {volume} {94}},\ \bibinfo {pages} {3642} (\bibinfo {year}
  {2014})}\BibitemShut {NoStop}%
\bibitem [{\citenamefont {Katayama}\ \emph {et~al.}(2009)\citenamefont
  {Katayama}, \citenamefont {Uchida}, \citenamefont {Hashizume}, \citenamefont
  {Niitaka}, \citenamefont {Matsuno}, \citenamefont {Matsumura}, \citenamefont
  {Nishihata}, \citenamefont {Mizuki}, \citenamefont {Takeshita}, \citenamefont
  {Gauzzi}, \citenamefont {Nohara},\ and\ \citenamefont {Takagi}}]{kat09}%
  \BibitemOpen
  \bibfield  {author} {\bibinfo {author} {\bibfnamefont {N.}~\bibnamefont
  {Katayama}}, \bibinfo {author} {\bibfnamefont {M.}~\bibnamefont {Uchida}},
  \bibinfo {author} {\bibfnamefont {D.}~\bibnamefont {Hashizume}}, \bibinfo
  {author} {\bibfnamefont {S.}~\bibnamefont {Niitaka}}, \bibinfo {author}
  {\bibfnamefont {J.}~\bibnamefont {Matsuno}}, \bibinfo {author} {\bibfnamefont
  {D.}~\bibnamefont {Matsumura}}, \bibinfo {author} {\bibfnamefont
  {Y.}~\bibnamefont {Nishihata}}, \bibinfo {author} {\bibfnamefont
  {J.}~\bibnamefont {Mizuki}}, \bibinfo {author} {\bibfnamefont
  {N.}~\bibnamefont {Takeshita}}, \bibinfo {author} {\bibfnamefont
  {A.}~\bibnamefont {Gauzzi}}, \bibinfo {author} {\bibfnamefont
  {M.}~\bibnamefont {Nohara}}, \ and\ \bibinfo {author} {\bibfnamefont
  {H.}~\bibnamefont {Takagi}},\ }\href {\doibase
  10.1103/PhysRevLett.103.146405} {\bibfield  {journal} {\bibinfo  {journal}
  {Phys. Rev. Lett.}\ }\textbf {\bibinfo {volume} {103}},\ \bibinfo {pages}
  {146405} (\bibinfo {year} {2009})}\BibitemShut {NoStop}%
\bibitem [{\citenamefont {Diep}(2005)}]{die05}%
  \BibitemOpen
  \bibfield  {author} {\bibinfo {author} {\bibfnamefont {H.~T.}\ \bibnamefont
  {Diep}},\ }\href@noop {} {\emph {\bibinfo {title} {Frustrated Spin
  Systems}}}\ (\bibinfo  {publisher} {World Scientific Publishing Company},\
  \bibinfo {year} {2005})\BibitemShut {NoStop}%
\bibitem [{\citenamefont {van Laar}\ and\ \citenamefont {Ijdo}(1971)}]{vla71}%
  \BibitemOpen
  \bibfield  {author} {\bibinfo {author} {\bibfnamefont {B.}~\bibnamefont {van
  Laar}}\ and\ \bibinfo {author} {\bibfnamefont {D.}~\bibnamefont {Ijdo}},\
  }\href {\doibase 10.1016/0022-4596(71)90106-X} {\bibfield  {journal}
  {\bibinfo  {journal} {Journal of Solid State Chemistry}\ }\textbf {\bibinfo
  {volume} {3}},\ \bibinfo {pages} {590 } (\bibinfo {year} {1971})}\BibitemShut
  {NoStop}%
\bibitem [{\citenamefont {Kawasaki}\ \emph {et~al.}(2011)\citenamefont
  {Kawasaki}, \citenamefont {Kishimoto}, \citenamefont {Tanaka}, \citenamefont
  {Ohno}, \citenamefont {Niitaka}, \citenamefont {Katayama},\ and\
  \citenamefont {Takagi}}]{kaw11}%
  \BibitemOpen
  \bibfield  {author} {\bibinfo {author} {\bibfnamefont {Y.}~\bibnamefont
  {Kawasaki}}, \bibinfo {author} {\bibfnamefont {Y.}~\bibnamefont {Kishimoto}},
  \bibinfo {author} {\bibfnamefont {T.}~\bibnamefont {Tanaka}}, \bibinfo
  {author} {\bibfnamefont {T.}~\bibnamefont {Ohno}}, \bibinfo {author}
  {\bibfnamefont {S.}~\bibnamefont {Niitaka}}, \bibinfo {author} {\bibfnamefont
  {N.}~\bibnamefont {Katayama}}, \ and\ \bibinfo {author} {\bibfnamefont
  {H.}~\bibnamefont {Takagi}},\ }\href@noop {} {\bibfield  {journal} {\bibinfo
  {journal} {Journal of Physics: Conference Series}\ }\textbf {\bibinfo
  {volume} {320}},\ \bibinfo {pages} {012028} (\bibinfo {year}
  {2011})}\BibitemShut {NoStop}%
\bibitem [{\citenamefont {Pen}\ \emph {et~al.}(1997{\natexlab{a}})\citenamefont
  {Pen}, \citenamefont {van~den Brink}, \citenamefont {Khomskii},\ and\
  \citenamefont {Sawatzky}}]{pen97}%
  \BibitemOpen
  \bibfield  {author} {\bibinfo {author} {\bibfnamefont {H.~F.}\ \bibnamefont
  {Pen}}, \bibinfo {author} {\bibfnamefont {J.}~\bibnamefont {van~den Brink}},
  \bibinfo {author} {\bibfnamefont {D.~I.}\ \bibnamefont {Khomskii}}, \ and\
  \bibinfo {author} {\bibfnamefont {G.~A.}\ \bibnamefont {Sawatzky}},\ }\href
  {\doibase 10.1103/PhysRevLett.78.1323} {\bibfield  {journal} {\bibinfo
  {journal} {Phys. Rev. Lett.}\ }\textbf {\bibinfo {volume} {78}},\ \bibinfo
  {pages} {1323} (\bibinfo {year} {1997}{\natexlab{a}})}\BibitemShut {NoStop}%
\bibitem [{\citenamefont {McQueen}\ \emph {et~al.}(2008)\citenamefont
  {McQueen}, \citenamefont {Stephens}, \citenamefont {Huang}, \citenamefont
  {Klimczuk}, \citenamefont {Ronning},\ and\ \citenamefont {Cava}}]{mcq08}%
  \BibitemOpen
  \bibfield  {author} {\bibinfo {author} {\bibfnamefont {T.~M.}\ \bibnamefont
  {McQueen}}, \bibinfo {author} {\bibfnamefont {P.~W.}\ \bibnamefont
  {Stephens}}, \bibinfo {author} {\bibfnamefont {Q.}~\bibnamefont {Huang}},
  \bibinfo {author} {\bibfnamefont {T.}~\bibnamefont {Klimczuk}}, \bibinfo
  {author} {\bibfnamefont {F.}~\bibnamefont {Ronning}}, \ and\ \bibinfo
  {author} {\bibfnamefont {R.~J.}\ \bibnamefont {Cava}},\ }\href {\doibase
  10.1103/PhysRevLett.101.166402} {\bibfield  {journal} {\bibinfo  {journal}
  {Phys. Rev. Lett.}\ }\textbf {\bibinfo {volume} {101}},\ \bibinfo {pages}
  {166402} (\bibinfo {year} {2008})}\BibitemShut {NoStop}%
\bibitem [{\citenamefont {Jin-no}\ \emph {et~al.}(2013)\citenamefont {Jin-no},
  \citenamefont {Shimizu}, \citenamefont {Itoh}, \citenamefont {Niitaka},\ and\
  \citenamefont {Takagi}}]{jin13}%
  \BibitemOpen
  \bibfield  {author} {\bibinfo {author} {\bibfnamefont {T.}~\bibnamefont
  {Jin-no}}, \bibinfo {author} {\bibfnamefont {Y.}~\bibnamefont {Shimizu}},
  \bibinfo {author} {\bibfnamefont {M.}~\bibnamefont {Itoh}}, \bibinfo {author}
  {\bibfnamefont {S.}~\bibnamefont {Niitaka}}, \ and\ \bibinfo {author}
  {\bibfnamefont {H.}~\bibnamefont {Takagi}},\ }\href {\doibase
  10.1103/PhysRevB.87.075135} {\bibfield  {journal} {\bibinfo  {journal} {Phys.
  Rev. B}\ }\textbf {\bibinfo {volume} {87}},\ \bibinfo {pages} {075135}
  (\bibinfo {year} {2013})}\BibitemShut {NoStop}%
\bibitem [{\citenamefont {Brouet}\ \emph {et~al.}(2013)\citenamefont {Brouet},
  \citenamefont {Mauchain}, \citenamefont {Papalazarou}, \citenamefont {Faure},
  \citenamefont {Marsi}, \citenamefont {Lin}, \citenamefont {Taleb-Ibrahimi},
  \citenamefont {Le~F\`evre}, \citenamefont {Bertran}, \citenamefont {Cario},
  \citenamefont {Janod}, \citenamefont {Corraze}, \citenamefont {Phuoc},\ and\
  \citenamefont {Perfetti}}]{bro13}%
  \BibitemOpen
  \bibfield  {author} {\bibinfo {author} {\bibfnamefont {V.}~\bibnamefont
  {Brouet}}, \bibinfo {author} {\bibfnamefont {J.}~\bibnamefont {Mauchain}},
  \bibinfo {author} {\bibfnamefont {E.}~\bibnamefont {Papalazarou}}, \bibinfo
  {author} {\bibfnamefont {J.}~\bibnamefont {Faure}}, \bibinfo {author}
  {\bibfnamefont {M.}~\bibnamefont {Marsi}}, \bibinfo {author} {\bibfnamefont
  {P.~H.}\ \bibnamefont {Lin}}, \bibinfo {author} {\bibfnamefont
  {A.}~\bibnamefont {Taleb-Ibrahimi}}, \bibinfo {author} {\bibfnamefont
  {P.}~\bibnamefont {Le~F\`evre}}, \bibinfo {author} {\bibfnamefont
  {F.}~\bibnamefont {Bertran}}, \bibinfo {author} {\bibfnamefont
  {L.}~\bibnamefont {Cario}}, \bibinfo {author} {\bibfnamefont
  {E.}~\bibnamefont {Janod}}, \bibinfo {author} {\bibfnamefont
  {B.}~\bibnamefont {Corraze}}, \bibinfo {author} {\bibfnamefont {V.~T.}\
  \bibnamefont {Phuoc}}, \ and\ \bibinfo {author} {\bibfnamefont
  {L.}~\bibnamefont {Perfetti}},\ }\href {\doibase 10.1103/PhysRevB.87.041106}
  {\bibfield  {journal} {\bibinfo  {journal} {Phys. Rev. B}\ }\textbf {\bibinfo
  {volume} {87}},\ \bibinfo {pages} {041106} (\bibinfo {year}
  {2013})}\BibitemShut {NoStop}%
\bibitem [{\citenamefont {Pen}\ \emph {et~al.}(1997{\natexlab{b}})\citenamefont
  {Pen}, \citenamefont {Tjeng}, \citenamefont {Pellegrin}, \citenamefont
  {de~Groot}, \citenamefont {Sawatzky}, \citenamefont {van Veenendaal},\ and\
  \citenamefont {Chen}}]{pen98}%
  \BibitemOpen
  \bibfield  {author} {\bibinfo {author} {\bibfnamefont {H.~F.}\ \bibnamefont
  {Pen}}, \bibinfo {author} {\bibfnamefont {L.~H.}\ \bibnamefont {Tjeng}},
  \bibinfo {author} {\bibfnamefont {E.}~\bibnamefont {Pellegrin}}, \bibinfo
  {author} {\bibfnamefont {F.~M.~F.}\ \bibnamefont {de~Groot}}, \bibinfo
  {author} {\bibfnamefont {G.~A.}\ \bibnamefont {Sawatzky}}, \bibinfo {author}
  {\bibfnamefont {M.~A.}\ \bibnamefont {van Veenendaal}}, \ and\ \bibinfo
  {author} {\bibfnamefont {C.~T.}\ \bibnamefont {Chen}},\ }\href {\doibase
  10.1103/PhysRevB.55.15500} {\bibfield  {journal} {\bibinfo  {journal} {Phys.
  Rev. B}\ }\textbf {\bibinfo {volume} {55}},\ \bibinfo {pages} {15500}
  (\bibinfo {year} {1997}{\natexlab{b}})}\BibitemShut {NoStop}%
\bibitem [{\citenamefont {Guo}\ \emph {et~al.}(2011)\citenamefont {Guo},
  \citenamefont {Zhang}, \citenamefont {Zhang}, \citenamefont {Jia},\ and\
  \citenamefont {Zeng}}]{guo11}%
  \BibitemOpen
  \bibfield  {author} {\bibinfo {author} {\bibfnamefont {Y.}~\bibnamefont
  {Guo}}, \bibinfo {author} {\bibfnamefont {G.}~\bibnamefont {Zhang}}, \bibinfo
  {author} {\bibfnamefont {X.}~\bibnamefont {Zhang}}, \bibinfo {author}
  {\bibfnamefont {T.}~\bibnamefont {Jia}}, \ and\ \bibinfo {author}
  {\bibfnamefont {Z.}~\bibnamefont {Zeng}},\ }\href {\doibase
  10.1063/1.3562452} {\bibfield  {journal} {\bibinfo  {journal} {Journal of
  Applied Physics}\ }\textbf {\bibinfo {volume} {109}},\ \bibinfo {eid}
  {07E145} (\bibinfo {year} {2011})}\BibitemShut {NoStop}%
\bibitem [{\citenamefont {Yoshitake}\ and\ \citenamefont
  {Motome}(2011)}]{yos11}%
  \BibitemOpen
  \bibfield  {author} {\bibinfo {author} {\bibfnamefont {J.}~\bibnamefont
  {Yoshitake}}\ and\ \bibinfo {author} {\bibfnamefont {Y.}~\bibnamefont
  {Motome}},\ }\href {\doibase 10.1143/JPSJ.80.073711} {\bibfield  {journal}
  {\bibinfo  {journal} {Journal of the Physical Society of Japan}\ }\textbf
  {\bibinfo {volume} {80}},\ \bibinfo {pages} {073711} (\bibinfo {year}
  {2011})}\BibitemShut {NoStop}%
\bibitem [{\citenamefont {Ezhov}\ \emph {et~al.}(1998)\citenamefont {Ezhov},
  \citenamefont {Anisimov}, \citenamefont {Pen}, \citenamefont {Khomskii},\
  and\ \citenamefont {Sawatzky}}]{ezh98}%
  \BibitemOpen
  \bibfield  {author} {\bibinfo {author} {\bibfnamefont {S.~Y.}\ \bibnamefont
  {Ezhov}}, \bibinfo {author} {\bibfnamefont {V.~I.}\ \bibnamefont {Anisimov}},
  \bibinfo {author} {\bibfnamefont {H.~F.}\ \bibnamefont {Pen}}, \bibinfo
  {author} {\bibfnamefont {D.~I.}\ \bibnamefont {Khomskii}}, \ and\ \bibinfo
  {author} {\bibfnamefont {G.~A.}\ \bibnamefont {Sawatzky}},\ }\href@noop {}
  {\bibfield  {journal} {\bibinfo  {journal} {EPL (Europhysics Letters)}\
  }\textbf {\bibinfo {volume} {44}},\ \bibinfo {pages} {491} (\bibinfo {year}
  {1998})}\BibitemShut {NoStop}%
\bibitem [{\citenamefont {Georges}\ \emph {et~al.}(1996)\citenamefont
  {Georges}, \citenamefont {Kotliar}, \citenamefont {Krauth},\ and\
  \citenamefont {Rozenberg}}]{geo96}%
  \BibitemOpen
  \bibfield  {author} {\bibinfo {author} {\bibfnamefont {A.}~\bibnamefont
  {Georges}}, \bibinfo {author} {\bibfnamefont {G.}~\bibnamefont {Kotliar}},
  \bibinfo {author} {\bibfnamefont {W.}~\bibnamefont {Krauth}}, \ and\ \bibinfo
  {author} {\bibfnamefont {M.~J.}\ \bibnamefont {Rozenberg}},\ }\href {\doibase
  10.1103/RevModPhys.68.13} {\bibfield  {journal} {\bibinfo  {journal} {Rev.
  Mod. Phys.}\ }\textbf {\bibinfo {volume} {68}},\ \bibinfo {pages} {13}
  (\bibinfo {year} {1996})}\BibitemShut {NoStop}%
\bibitem [{\citenamefont {Anisimov}\ \emph {et~al.}(1997)\citenamefont
  {Anisimov}, \citenamefont {Poteryaev}, \citenamefont {Korotin}, \citenamefont
  {Anokhin},\ and\ \citenamefont {Kotliar}}]{ani97}%
  \BibitemOpen
  \bibfield  {author} {\bibinfo {author} {\bibfnamefont {V.~I.}\ \bibnamefont
  {Anisimov}}, \bibinfo {author} {\bibfnamefont {A.~I.}\ \bibnamefont
  {Poteryaev}}, \bibinfo {author} {\bibfnamefont {M.~A.}\ \bibnamefont
  {Korotin}}, \bibinfo {author} {\bibfnamefont {A.~O.}\ \bibnamefont
  {Anokhin}}, \ and\ \bibinfo {author} {\bibfnamefont {G.}~\bibnamefont
  {Kotliar}},\ }\href@noop {} {\bibfield  {journal} {\bibinfo  {journal}
  {Journal of Physics: Condensed Matter}\ }\textbf {\bibinfo {volume} {9}},\
  \bibinfo {pages} {7359} (\bibinfo {year} {1997})}\BibitemShut {NoStop}%
\bibitem [{\citenamefont {Lichtenstein}\ and\ \citenamefont
  {Katsnelson}(1998)}]{lic98}%
  \BibitemOpen
  \bibfield  {author} {\bibinfo {author} {\bibfnamefont {A.~I.}\ \bibnamefont
  {Lichtenstein}}\ and\ \bibinfo {author} {\bibfnamefont {M.~I.}\ \bibnamefont
  {Katsnelson}},\ }\href {\doibase 10.1103/PhysRevB.57.6884} {\bibfield
  {journal} {\bibinfo  {journal} {Phys. Rev. B}\ }\textbf {\bibinfo {volume}
  {57}},\ \bibinfo {pages} {6884} (\bibinfo {year} {1998})}\BibitemShut
  {NoStop}%
\bibitem [{\citenamefont {Kotliar}\ \emph {et~al.}(2006)\citenamefont
  {Kotliar}, \citenamefont {Savrasov}, \citenamefont {Haule}, \citenamefont
  {Oudovenko}, \citenamefont {Parcollet},\ and\ \citenamefont
  {Marianetti}}]{kot06}%
  \BibitemOpen
  \bibfield  {author} {\bibinfo {author} {\bibfnamefont {G.}~\bibnamefont
  {Kotliar}}, \bibinfo {author} {\bibfnamefont {S.~Y.}\ \bibnamefont
  {Savrasov}}, \bibinfo {author} {\bibfnamefont {K.}~\bibnamefont {Haule}},
  \bibinfo {author} {\bibfnamefont {V.~S.}\ \bibnamefont {Oudovenko}}, \bibinfo
  {author} {\bibfnamefont {O.}~\bibnamefont {Parcollet}}, \ and\ \bibinfo
  {author} {\bibfnamefont {C.~A.}\ \bibnamefont {Marianetti}},\ }\href
  {\doibase 10.1103/RevModPhys.78.865} {\bibfield  {journal} {\bibinfo
  {journal} {Rev. Mod. Phys.}\ }\textbf {\bibinfo {volume} {78}},\ \bibinfo
  {pages} {865} (\bibinfo {year} {2006})}\BibitemShut {NoStop}%
\bibitem [{\citenamefont {Maier}\ \emph {et~al.}(2005)\citenamefont {Maier},
  \citenamefont {Jarrell}, \citenamefont {Pruschke},\ and\ \citenamefont
  {Hettler}}]{mai05}%
  \BibitemOpen
  \bibfield  {author} {\bibinfo {author} {\bibfnamefont {T.}~\bibnamefont
  {Maier}}, \bibinfo {author} {\bibfnamefont {M.}~\bibnamefont {Jarrell}},
  \bibinfo {author} {\bibfnamefont {T.}~\bibnamefont {Pruschke}}, \ and\
  \bibinfo {author} {\bibfnamefont {M.~H.}\ \bibnamefont {Hettler}},\ }\href
  {\doibase 10.1103/RevModPhys.77.1027} {\bibfield  {journal} {\bibinfo
  {journal} {Rev. Mod. Phys.}\ }\textbf {\bibinfo {volume} {77}},\ \bibinfo
  {pages} {1027} (\bibinfo {year} {2005})}\BibitemShut {NoStop}%
\bibitem [{\citenamefont {Boehnke}\ and\ \citenamefont
  {Lechermann}(2012)}]{boe12}%
  \BibitemOpen
  \bibfield  {author} {\bibinfo {author} {\bibfnamefont {L.}~\bibnamefont
  {Boehnke}}\ and\ \bibinfo {author} {\bibfnamefont {F.}~\bibnamefont
  {Lechermann}},\ }\href {\doibase 10.1103/PhysRevB.85.115128} {\bibfield
  {journal} {\bibinfo  {journal} {Phys. Rev. B}\ }\textbf {\bibinfo {volume}
  {85}},\ \bibinfo {pages} {115128} (\bibinfo {year} {2012})}\BibitemShut
  {NoStop}%
\bibitem [{\citenamefont {Boehnke}\ \emph {et~al.}(2011)\citenamefont
  {Boehnke}, \citenamefont {Hafermann}, \citenamefont {Ferrero}, \citenamefont
  {Lechermann},\ and\ \citenamefont {Parcollet}}]{boe11}%
  \BibitemOpen
  \bibfield  {author} {\bibinfo {author} {\bibfnamefont {L.}~\bibnamefont
  {Boehnke}}, \bibinfo {author} {\bibfnamefont {H.}~\bibnamefont {Hafermann}},
  \bibinfo {author} {\bibfnamefont {M.}~\bibnamefont {Ferrero}}, \bibinfo
  {author} {\bibfnamefont {F.}~\bibnamefont {Lechermann}}, \ and\ \bibinfo
  {author} {\bibfnamefont {O.}~\bibnamefont {Parcollet}},\ }\href {\doibase
  10.1103/PhysRevB.84.075145} {\bibfield  {journal} {\bibinfo  {journal} {Phys.
  Rev. B}\ }\textbf {\bibinfo {volume} {84}},\ \bibinfo {pages} {075145}
  (\bibinfo {year} {2011})}\BibitemShut {NoStop}%
\bibitem [{\citenamefont {Boehnke}(2015)}]{boe15}%
  \BibitemOpen
  \bibfield  {author} {\bibinfo {author} {\bibfnamefont {L.}~\bibnamefont
  {Boehnke}},\ }\emph {\bibinfo {title} {Susceptibilities in materials with
  multiple strongly correlated orbitals}},\ \href
  {http://ediss.sub.uni-hamburg.de/volltexte/2015/7325} {Ph.D. thesis},\
  \bibinfo  {school} {Universit{\"a}t Hamburg} (\bibinfo {year}
  {2015})\BibitemShut {NoStop}%
\bibitem [{\citenamefont {Pavarini}\ \emph {et~al.}(2008)\citenamefont
  {Pavarini}, \citenamefont {Koch},\ and\ \citenamefont
  {Lichtenstein}}]{pav08}%
  \BibitemOpen
  \bibfield  {author} {\bibinfo {author} {\bibfnamefont {E.}~\bibnamefont
  {Pavarini}}, \bibinfo {author} {\bibfnamefont {E.}~\bibnamefont {Koch}}, \
  and\ \bibinfo {author} {\bibfnamefont {A.~I.}\ \bibnamefont {Lichtenstein}},\
  }\href {\doibase 10.1103/PhysRevLett.101.266405} {\bibfield  {journal}
  {\bibinfo  {journal} {Phys. Rev. Lett.}\ }\textbf {\bibinfo {volume} {101}},\
  \bibinfo {pages} {266405} (\bibinfo {year} {2008})}\BibitemShut {NoStop}%
\bibitem [{\citenamefont {Meyer}\ \emph {et~al.}()\citenamefont {Meyer},
  \citenamefont {Els\"asser}, \citenamefont {Lechermann},\ and\ \citenamefont
  {F\"ahnle}}]{mbpp}%
  \BibitemOpen
  \bibfield  {author} {\bibinfo {author} {\bibfnamefont {B.}~\bibnamefont
  {Meyer}}, \bibinfo {author} {\bibfnamefont {C.}~\bibnamefont {Els\"asser}},
  \bibinfo {author} {\bibfnamefont {F.}~\bibnamefont {Lechermann}}, \ and\
  \bibinfo {author} {\bibfnamefont {M.}~\bibnamefont {F\"ahnle}},\ }\href@noop
  {} {\enquote {\bibinfo {title} {{FORTAN 90 Program for
  Mixed-Basis-Pseudopotential Calculations for Crystals}},}\ }\bibinfo {note}
  {{Max-Planck-Institut f\"ur Metallforschung, Stuttgart}}\BibitemShut
  {NoStop}%
\bibitem [{\citenamefont {Marzari}\ and\ \citenamefont
  {Vanderbilt}(1997)}]{mar97}%
  \BibitemOpen
  \bibfield  {author} {\bibinfo {author} {\bibfnamefont {N.}~\bibnamefont
  {Marzari}}\ and\ \bibinfo {author} {\bibfnamefont {D.}~\bibnamefont
  {Vanderbilt}},\ }\href {\doibase 10.1103/PhysRevB.56.12847} {\bibfield
  {journal} {\bibinfo  {journal} {Phys. Rev. B}\ }\textbf {\bibinfo {volume}
  {56}},\ \bibinfo {pages} {12847} (\bibinfo {year} {1997})}\BibitemShut
  {NoStop}%
\bibitem [{\citenamefont {Mostofi}\ \emph {et~al.}(2008)\citenamefont
  {Mostofi}, \citenamefont {Yates}, \citenamefont {Lee}, \citenamefont {Souza},
  \citenamefont {Vanderbilt},\ and\ \citenamefont {Marzari}}]{mos08}%
  \BibitemOpen
  \bibfield  {author} {\bibinfo {author} {\bibfnamefont {A.~A.}\ \bibnamefont
  {Mostofi}}, \bibinfo {author} {\bibfnamefont {J.~R.}\ \bibnamefont {Yates}},
  \bibinfo {author} {\bibfnamefont {Y.-S.}\ \bibnamefont {Lee}}, \bibinfo
  {author} {\bibfnamefont {I.}~\bibnamefont {Souza}}, \bibinfo {author}
  {\bibfnamefont {D.}~\bibnamefont {Vanderbilt}}, \ and\ \bibinfo {author}
  {\bibfnamefont {N.}~\bibnamefont {Marzari}},\ }\href {\doibase
  10.1016/j.cpc.2007.11.016} {\bibfield  {journal} {\bibinfo  {journal}
  {Computer Physics Communications}\ }\textbf {\bibinfo {volume} {178}},\
  \bibinfo {pages} {685 } (\bibinfo {year} {2008})}\BibitemShut {NoStop}%
\bibitem [{\citenamefont {Georges}\ \emph {et~al.}(2013)\citenamefont
  {Georges}, \citenamefont {Medici},\ and\ \citenamefont {Mravlje}}]{geo13}%
  \BibitemOpen
  \bibfield  {author} {\bibinfo {author} {\bibfnamefont {A.}~\bibnamefont
  {Georges}}, \bibinfo {author} {\bibfnamefont {L.~d.}\ \bibnamefont {Medici}},
  \ and\ \bibinfo {author} {\bibfnamefont {J.}~\bibnamefont {Mravlje}},\ }\href
  {\doibase 10.1146/annurev-conmatphys-020911-125045} {\bibfield  {journal}
  {\bibinfo  {journal} {Annual Review of Condensed Matter Physics}\ }\textbf
  {\bibinfo {volume} {4}},\ \bibinfo {pages} {137} (\bibinfo {year}
  {2013})}\BibitemShut {NoStop}%
\bibitem [{\citenamefont {Werner}\ \emph {et~al.}(2006)\citenamefont {Werner},
  \citenamefont {Comanac}, \citenamefont {de' Medici}, \citenamefont {Troyer},\
  and\ \citenamefont {Millis}}]{wer06}%
  \BibitemOpen
  \bibfield  {author} {\bibinfo {author} {\bibfnamefont {P.}~\bibnamefont
  {Werner}}, \bibinfo {author} {\bibfnamefont {A.}~\bibnamefont {Comanac}},
  \bibinfo {author} {\bibfnamefont {L.}~\bibnamefont {de' Medici}}, \bibinfo
  {author} {\bibfnamefont {M.}~\bibnamefont {Troyer}}, \ and\ \bibinfo {author}
  {\bibfnamefont {A.~J.}\ \bibnamefont {Millis}},\ }\href {\doibase
  10.1103/PhysRevLett.97.076405} {\bibfield  {journal} {\bibinfo  {journal}
  {Phys. Rev. Lett.}\ }\textbf {\bibinfo {volume} {97}},\ \bibinfo {pages}
  {076405} (\bibinfo {year} {2006})}\BibitemShut {NoStop}%
\bibitem [{\citenamefont {Gull}\ \emph {et~al.}(2011)\citenamefont {Gull},
  \citenamefont {Millis}, \citenamefont {Lichtenstein}, \citenamefont
  {Rubtsov}, \citenamefont {Troyer},\ and\ \citenamefont {Werner}}]{gul11}%
  \BibitemOpen
  \bibfield  {author} {\bibinfo {author} {\bibfnamefont {E.}~\bibnamefont
  {Gull}}, \bibinfo {author} {\bibfnamefont {A.~J.}\ \bibnamefont {Millis}},
  \bibinfo {author} {\bibfnamefont {A.~I.}\ \bibnamefont {Lichtenstein}},
  \bibinfo {author} {\bibfnamefont {A.~N.}\ \bibnamefont {Rubtsov}}, \bibinfo
  {author} {\bibfnamefont {M.}~\bibnamefont {Troyer}}, \ and\ \bibinfo {author}
  {\bibfnamefont {P.}~\bibnamefont {Werner}},\ }\href {\doibase
  10.1103/RevModPhys.83.349} {\bibfield  {journal} {\bibinfo  {journal} {Rev.
  Mod. Phys.}\ }\textbf {\bibinfo {volume} {83}},\ \bibinfo {pages} {349}
  (\bibinfo {year} {2011})}\BibitemShut {NoStop}%
\bibitem [{\citenamefont {Parcollet}\ \emph {et~al.}(2015)\citenamefont
  {Parcollet}, \citenamefont {Ferrero}, \citenamefont {Ayral}, \citenamefont
  {Hafermann}, \citenamefont {Krivenko}, \citenamefont {Messio},\ and\
  \citenamefont {Seth}}]{par15}%
  \BibitemOpen
  \bibfield  {author} {\bibinfo {author} {\bibfnamefont {O.}~\bibnamefont
  {Parcollet}}, \bibinfo {author} {\bibfnamefont {M.}~\bibnamefont {Ferrero}},
  \bibinfo {author} {\bibfnamefont {T.}~\bibnamefont {Ayral}}, \bibinfo
  {author} {\bibfnamefont {H.}~\bibnamefont {Hafermann}}, \bibinfo {author}
  {\bibfnamefont {I.}~\bibnamefont {Krivenko}}, \bibinfo {author}
  {\bibfnamefont {L.}~\bibnamefont {Messio}}, \ and\ \bibinfo {author}
  {\bibfnamefont {P.}~\bibnamefont {Seth}},\ }\href {\doibase
  http://dx.doi.org/10.1016/j.cpc.2015.04.023} {\bibfield  {journal} {\bibinfo
  {journal} {Computer Physics Communications}\ }\textbf {\bibinfo {volume}
  {196}},\ \bibinfo {pages} {398 } (\bibinfo {year} {2015})}\BibitemShut
  {NoStop}%
\bibitem [{\citenamefont {{Seth}}\ \emph {et~al.}(2015)\citenamefont {{Seth}},
  \citenamefont {{Krivenko}}, \citenamefont {{Ferrero}},\ and\ \citenamefont
  {{Parcollet}}}]{set15}%
  \BibitemOpen
  \bibfield  {author} {\bibinfo {author} {\bibfnamefont {P.}~\bibnamefont
  {{Seth}}}, \bibinfo {author} {\bibfnamefont {I.}~\bibnamefont {{Krivenko}}},
  \bibinfo {author} {\bibfnamefont {M.}~\bibnamefont {{Ferrero}}}, \ and\
  \bibinfo {author} {\bibfnamefont {O.}~\bibnamefont {{Parcollet}}},\
  }\href@noop {} {\bibfield  {journal} {\bibinfo  {journal} {ArXiv e-prints}\ }
  (\bibinfo {year} {2015})}\BibitemShut {NoStop}%
\bibitem [{\citenamefont {Strand}(2014)}]{str14}%
  \BibitemOpen
  \bibfield  {author} {\bibinfo {author} {\bibfnamefont {H.~U.~R.}\
  \bibnamefont {Strand}},\ }\href {\doibase 10.1103/PhysRevB.90.155108}
  {\bibfield  {journal} {\bibinfo  {journal} {Phys. Rev. B}\ }\textbf {\bibinfo
  {volume} {90}},\ \bibinfo {pages} {155108} (\bibinfo {year}
  {2014})}\BibitemShut {NoStop}%
\bibitem [{\citenamefont {de' Medici}\ \emph {et~al.}(2011)\citenamefont {de'
  Medici}, \citenamefont {Mravlje},\ and\ \citenamefont {Georges}}]{dem11}%
  \BibitemOpen
  \bibfield  {author} {\bibinfo {author} {\bibfnamefont {L.}~\bibnamefont {de'
  Medici}}, \bibinfo {author} {\bibfnamefont {J.}~\bibnamefont {Mravlje}}, \
  and\ \bibinfo {author} {\bibfnamefont {A.}~\bibnamefont {Georges}},\ }\href
  {\doibase 10.1103/PhysRevLett.107.256401} {\bibfield  {journal} {\bibinfo
  {journal} {Phys. Rev. Lett.}\ }\textbf {\bibinfo {volume} {107}},\ \bibinfo
  {pages} {256401} (\bibinfo {year} {2011})}\BibitemShut {NoStop}%
\bibitem [{\citenamefont {Grieger}\ \emph {et~al.}(2010)\citenamefont
  {Grieger}, \citenamefont {Boehnke},\ and\ \citenamefont
  {Lechermann}}]{gri10}%
  \BibitemOpen
  \bibfield  {author} {\bibinfo {author} {\bibfnamefont {D.}~\bibnamefont
  {Grieger}}, \bibinfo {author} {\bibfnamefont {L.}~\bibnamefont {Boehnke}}, \
  and\ \bibinfo {author} {\bibfnamefont {F.}~\bibnamefont {Lechermann}},\
  }\href@noop {} {\bibfield  {journal} {\bibinfo  {journal} {Journal of
  Physics: Condensed Matter}\ }\textbf {\bibinfo {volume} {22}},\ \bibinfo
  {pages} {275601} (\bibinfo {year} {2010})}\BibitemShut {NoStop}%
\bibitem [{\citenamefont {Zlatic}\ and\ \citenamefont
  {Horvatic}(1990)}]{zla90}%
  \BibitemOpen
  \bibfield  {author} {\bibinfo {author} {\bibfnamefont {V.}~\bibnamefont
  {Zlatic}}\ and\ \bibinfo {author} {\bibfnamefont {B.}~\bibnamefont
  {Horvatic}},\ }\href {\doibase 10.1016/0038-1098(90)90282-G} {\bibfield
  {journal} {\bibinfo  {journal} {Solid State Communications}\ }\textbf
  {\bibinfo {volume} {75}},\ \bibinfo {pages} {263 } (\bibinfo {year}
  {1990})}\BibitemShut {NoStop}%
\end{thebibliography}%

\end{document}